\documentclass{article}

\PassOptionsToPackage{numbers, compress}{natbib}
\usepackage[preprint]{neurips_2024}

\usepackage[utf8]{inputenc}
\usepackage[T1]{fontenc}
\usepackage{microtype}
\usepackage{amsmath, amssymb}
\usepackage{booktabs}
\usepackage{array}
\usepackage{tabularx}
\usepackage{graphicx}
\usepackage{xcolor}
\usepackage{hyperref}
\usepackage{url}

\title{Divergent Recommendations, Convergent Diagnoses:\\Cross-Provider Failure-Mode Convergence\\in AI Commercial Recommendation}

\author{%
  Will Jack\thanks{Equal contribution.} \quad
  Noah Lehman\footnotemark[1] \quad
  Keller Maloney\footnotemark[1] \quad
  Sarah Xu\footnotemark[1] \\
  Unusual \\
  \texttt{\{will, noah, keller, sarah\}@unusual.ai}
}

\date{May 21, 2026}

\begin{document}

\renewcommand{\thefootnote}{\fnsymbol{footnote}}
\maketitle
\renewcommand{\thefootnote}{\arabic{footnote}}
\setcounter{footnote}{0}

\begin{abstract}
\noindent
A brand whose customers use both ChatGPT and Claude for product recommendations faces a strategic choice: a single optimization playbook, or one per provider? Across 215 commercially-framed prompts in four measurement batches, the two providers disagree on which brands they recommend roughly two-thirds of the time (cross-provider recommendation Jaccard 0.35, below the 0.50--0.61 same-prompt rerun baseline). The picks diverge. But when neither provider recommends a brand, we classify the failure into one of three modes --- \emph{discoverability} (the brand never reaches the model), \emph{compellingness} (it reaches the model but isn't mentioned), or \emph{positioning} (it's mentioned but not recommended) --- and on 7{,}763 such joint failures, both providers diagnose the same failure mode 95.1\% of the time (clustered 95\% CI [94.3\%, 95.7\%]). Agreement rises monotonically with falling brand prominence, from 81\% [78.2\%, 84.0\%] on category leaders to 99.6\% [99.3\%, 99.9\%] on long-tail regional brands. The two providers reach their picks by measurably different generative routes --- Anthropic recommends from priors 43--52\% of the time, OpenAI 8--29\% --- but they converge on the failure diagnosis where it matters most for the long tail. Work that addresses the diagnosed failure mode lifts visibility on both providers; positioning- and content-level work for category leaders is more provider-specific.
\end{abstract}

\section{Introduction}

Brand audiences are now split across AI providers. Some buyers default to ChatGPT, others to Claude, and an increasing share to Gemini or Perplexity. A brand whose customers are distributed across multiple providers has to decide whether the same marketing-to-AI playbook serves all of them, or whether each provider requires its own. The financial stakes on that question are substantial: at scale, ``one plan per provider'' multiplies content production, measurement, and authority-list work; ``one plan transfers'' lets the same investment compound across surfaces. We treat ``marketing to AI'' as the umbrella problem --- positioning, content, product fit, and discoverability --- of which answer-engine and generative-engine optimization (AEO / GEO) is the narrower, discoverability-focused sub-practice.

The AEO / GEO tracking industry has organized around per-provider dashboards --- Profound, Otterly, LLM Pulse, HubSpot AEO Grader, Authoritas, Semrush, and others report visibility separately on ChatGPT, Claude, Perplexity, and Gemini, on the implicit premise that the \emph{picks} differ across providers. What's much less settled --- and what brand-side marketing teams actually need an answer to --- is whether the brand-side \emph{interventions} also differ: does work that helps a brand on one provider transfer to the others, or does each surface require its own content / authority / positioning playbook? \citet{yang2025} document substantial cross-provider divergence in news-source citation patterns alongside similar political-lean biases, consistent with the divergent-picks half of the question; \citet{aggarwal2024}'s GEO benchmark reports headline lift on a single engine (GPT-3.5-turbo) with separate, smaller numbers on Perplexity; the cross-provider transfer question is not the primary axis of that work. The most theoretically explicit prior --- \citet{bommasani2022}'s algorithmic-monoculture hypothesis --- predicts a sharper structural pattern that bears on both halves: providers that share training-corpus and architectural lineage should produce correlated \emph{exclusions} (which brands they fail to retrieve and recommend) alongside uncorrelated \emph{inclusions} (which brands they specifically pick from those that are retrieved).

We empirically separate the two layers --- picks and diagnoses --- and find precisely the asymmetry Bommasani et al.'s hypothesis predicts. The recommendation sets disagree more than within-cell reruns do, replicating across four independent measurement batches. The diagnostic-stage classification of \emph{joint failures} agrees overwhelmingly: 95.1\% of cells lie on the diagonal of the cross-provider funnel-stage confusion matrix (prompt-clustered 95\% CI: [94.3\%, 95.7\%]), and that figure rises with falling brand prominence to 99.6\% [99.3\%, 99.9\%] at L5.

The mechanism is also asymmetric in a way that adds texture rather than contradiction. Anthropic flagships recommend brands without any retrieval-layer evidence at 4--6$\times$ the rate of OpenAI cells (the gpt-5.4 flagship sits at the floor of this range; gpt-5.4-mini at the upper end); OpenAI's recommendation route is more retrieval-anchored. The systems arrive at differing picks via different generative routes, but when both providers fail to recommend a brand, they fail at the same stage --- typically because both have failed to discover it.

\subsection{Contributions}

This paper makes three contributions.

\begin{enumerate}
\item \textbf{Cross-provider recommendation-set divergence with confidence intervals.} Across four independent measurement batches, cross-provider per-prompt recommendation Jaccard sits at 0.349--0.356 with overlapping 95\% confidence intervals --- replicating cleanly, and falling below the within-cell rerun-stability baseline of 0.50--0.61. Provider choice changes the picks.

\item \textbf{Mechanistic asymmetry between priors-heavy and retrieval-heavy recommendation routes.} A pure-priors signal --- brands recommended without appearance in any retrieval-evidence layer --- separates the providers without Wilson 95\% CI overlap: Anthropic flagships at 43--52\%, OpenAI cells at 8--29\% (gpt-5.4 flagship 8--12\%; gpt-5.4-mini 21--29\%). OpenAI batch-reconstruction calibration corroborates the retrieval-anchored route: 60--71\% of brand-disambiguated queries appear in the very first search batch. The two providers reach their picks by different routes.

\item \textbf{Failure-mode convergence with a monotone prominence gradient.} On a 3$\times$3 funnel-stage confusion matrix restricted to 7{,}763 joint-failure cells, same-stage agreement is 95.1\% overall (prompt-clustered 95\% CI: [94.3\%, 95.7\%]; 1{,}000-iteration bootstrap resampling prompts with replacement). Cohen's $\kappa$ of 0.29 sits in the ``fair'' band of the Landis \& Koch (1977) classification above a high (93.0\%) chance-agreement baseline driven by Stage-1 dominance; per Section~6.4, restricted to cells where at least one provider's modal stage is S2 or S3 ($n{=}450$), diagonal agreement is 14.9\% --- so the pooled 95.1\% is an S1-convergence figure. The load-bearing inferential signal is the monotone gradient of same-stage agreement across brand prominence: 81.2\% [78.2\%, 84.0\%] at L1, 88.2\% [85.6\%, 90.8\%] at L2, 96.2\% [95.1\%, 97.2\%] at L3, 99.3\% [98.9\%, 99.7\%] at L4, 99.6\% [99.3\%, 99.9\%] at L5 (clustered 95\% CIs). The L1 CI excludes the pooled 95.1\% on the downside and the L5 CI excludes it on the upside; the gradient is not a Stage-1-dominance artifact. The diagnostic prescription converges where the recommendation does not.
\end{enumerate}

\section{Background}

Three lines of prior work bear directly on the dichotomy we measure.

\textbf{Cross-provider LLM comparison.} The HELM framework \citep{liang2023} established the methodological template of evaluating multiple providers under matched conditions across a battery of standardized tasks. Subsequent commercial-domain work has documented systematic cross-provider behavioral differences: \citet{yang2025}, analyzing 366K citations across OpenAI, Perplexity, and Google, reports pronounced inter-provider Gini differences in citation concentration alongside similar political-lean biases --- the same divergent-surface / convergent-bias pattern we observe in commercial recommendation. Our audit extends the HELM-lineage cross-provider comparison from static QA to live, search-augmented commercial recommendation, and decomposes the cross-provider behavior into a separable pick layer and diagnosis layer.

\textbf{Priors versus retrieval in retrieval-augmented LLMs.} \citet{mallen2023} showed that language models answer popular entities from parametric memory but require retrieval for long-tail entities --- a popularity-stratified parametric/non-parametric balance. \citet{xu2024} surveys the broader landscape of knowledge conflicts in LLMs and notes that no definitive rule yet exists for whether models prioritize contextual or parametric knowledge, with disparities attributed to training-data and post-training variations. \citet{goyal2025} document a counterintuitive ``context-parametric inversion'' where instruction-tuned models often \emph{increase} parametric reliance under conflicting context. Together these results provide the mechanistic vocabulary for our pure-priors signal: when a provider recommends a brand without any retrieval-layer evidence for it, the provider is operating on parametric memory, and the cross-provider rate of doing so is a measurable behavior.

\textbf{Algorithmic monoculture and the convergent-failure prediction.} \citet{bommasani2022} formalize the component-sharing hypothesis: deployments that share training data and architectural choices should produce correlated outcomes in aggregate. Translated to our setting, the prediction is that providers should be correlated in which brands they exclude (because the underlying retrieval corpora, the web, and the consensus authority lists are shared substrate) while being uncorrelated in which brands they specifically promote (because post-training and retrieval idiosyncrasy diverge). The 95.1\% diagnostic convergence we report is the first concrete commercial-recommendation evidence we know of for this prediction.

We adopt two definitional frameworks from \citet{jack2026prominence}, the companion prominence-stratified audit. First, a four-stage funnel taxonomy --- Stage 1 (S1, discoverability: the brand never reaches the model), Stage 2 (S2, compellingness: it reaches but isn't mentioned), Stage 3 (S3, positioning: it's mentioned but not recommended), Stage 4 (the success state) --- itself a recommendation-domain instantiation of the consumer decision journey \citep{court2009}. Funnel-stage failure modes admit a recommendation-specific reading along an omission-versus-content-error axis. The hallucination-taxonomy survey by \citet{cossio2025} organizes hallucinations along intrinsic-vs-extrinsic and factuality-vs-faithfulness axes; the omission-vs-content-error reading we adopt for the funnel is our own simplification rather than the survey's primary distinction. Same-stage agreement on a joint-failure cell is convergence on omission stage, not on answer content. Second, a five-tier brand-prominence ladder --- L1 category leaders, L2 established challengers, L3 mid-market, L4 long-tail specialists, L5 regional players --- assigned by \citet{jack2026prominence} to each of 533 reference brands. We use L1--L5 throughout to refer to the prominence label of a brand.

\section{Method}

\subsection{Cross-provider pairing}

We compare two provider pools: the OpenAI pool, comprising \texttt{gpt-5.4-mini} at low and high reasoning effort; and the Anthropic pool, comprising \texttt{claude-sonnet-4-6} at low and high reasoning effort. Within-pool aggregation is intentional: we report cross-provider effects net of reasoning-effort variation within each provider rather than fully crossing the effort axis with the provider axis. Where finer flagship-vs-flagship structure matters (mainly for the pure-priors signal) we report \texttt{gpt-5.4} and \texttt{claude-opus-4-6} cells separately and flag them as ancillary.

Prompts and runs are held constant across providers. For each prompt the OpenAI pool and the Anthropic pool issue search-augmented runs against the same prompt corpus, the same shadow-corpus reference brands (the 533-brand prominence-labeled catalog from \citealt{jack2026prominence}), and the same cross-judge consensus extraction pipeline. The four measurement batches used for the Jaccard analysis below correspond to four independent runs of the cross-provider design at differing scale.

\subsection{Cross-judge consensus extraction}

Following \citet{jack2026prominence}, brand mentions in completion text, snippets, titles, and issued search queries are extracted by two LLM judges (\texttt{claude-haiku-4-5} at low effort and \texttt{gpt-5-mini}) operating in consensus mode --- a brand is counted as present in a slot if and only if both judges identify it. The consensus protocol is a deliberate conservative choice that absorbs single-judge instability \citep{zheng2023}. It under-counts mentions at the margins but provides cross-provider robustness for downstream comparisons; in particular, both providers are evaluated by the same two judges.

\subsection{Funnel-stage classification}

For each (run $\times$ shadow-corpus brand) cell where the brand is sector-relevant to the prompt, we classify the brand's terminal funnel stage following the four-stage scheme introduced in \citet{jack2026prominence}:

\begin{itemize}
\item \textbf{Stage 1 (S1) --- no retrieval, no mention}: brand absent from every retrieval layer (issued queries, retrieved domains, snippet texts, snippet titles) \emph{and} from the completion.
\item \textbf{Stage 2 (S2) --- retrieval, no mention}: brand present in at least one retrieval layer but absent from the completion.
\item \textbf{Stage 3 (S3) --- mention, not recommended}: brand present in the completion but not consensus-classified as \texttt{recommended} by both judges.
\item \textbf{Stage 4 (S4) --- recommended}: brand consensus-recommended.
\end{itemize}

Stages 1--3 are mutually exclusive failure modes for the same brand under the same prompt; Stage 4 is the success state. The four labels are descriptive of \emph{what was observed in the run record} (presence/absence in retrieval, presence/absence in completion, presence/absence of consensus recommendation) rather than of cause. We interpret S1 as a discoverability failure, S2 as a compellingness failure, and S3 as a positioning failure --- against the broader hallucination-taxonomy literature surveyed by \citet{cossio2025}, we further read S1 as an omission-style failure and S2--S3 as content-error-style failures within a generated answer --- but the stage classification itself does not commit to those readings. A brand at S2 may be at S2 because it lost relevance, because list-length truncated it, because the persona did not match, or for other reasons we do not separately isolate.

\subsection{Modal-stage aggregation and joint-failure restriction}

For each (prompt $\times$ brand $\times$ provider) cell we take the modal terminal stage across the runs at that cell, requiring at least three runs per cell to assign a modal stage. The modal-stage operation absorbs within-cell run-to-run noise that, on its own, would inflate cross-provider disagreement.

We then restrict to \emph{joint-failure cells}: (prompt $\times$ brand) pairs where neither provider's modal stage equals Stage 4. The restriction is essential to the interpretation. The relevant scientific question is not ``do the providers agree on when they recommend a brand'' (a confounded question, since brands are sometimes simply universally recommended) but ``when neither recommends, do they agree on why.'' The joint-failure restriction isolates the diagnostic question.

The cross-provider failure-mode analysis uses 7{,}763 joint-failure cells. The marginal stage distributions are heavily skewed toward Stage 1 (7,512 cells for OpenAI; 7,457 for Anthropic), reflecting that the most common reason both providers fail to recommend a brand is that neither has discovered it.

\subsection{Pure-priors signal}

We define the \emph{pure-priors share} for a (cell) as the proportion of recommended-brand events in which the brand appears in the completion as recommended but does not appear in any retrieval-evidence layer for that run --- not in any issued query token, not in any retrieved domain, not in any snippet text, not in any snippet title. The conservative pure-priors definition does not require us to identify which retrieved page produced parametric versus retrieval recall; it requires only that the brand is absent from every retrieval-layer artifact we can observe.

The pure-priors share is more precisely a \emph{retrieval-unattributed} recommendation share: it counts recommended-brand events for which we observe no retrieval-layer evidence (in issued queries, retrieved domains, snippet text, or snippet titles). It is a one-sided signal. A cell with a low pure-priors share does not establish that the provider is purely retrieval-driven; recommended brands can appear in retrieval layers via incidental mentions independent of the recall pathway. A cell with a high pure-priors share is consistent with a parametric-memory recommendation route, but does not strictly establish it: fetched page text we do not log, brand aliases not normalized, brand recall mediated by a retrieved category page rather than a brand-specific snippet, and other hidden retrieval pathways would all show up as ``no observed retrieval-layer evidence.'' We treat parametric reliance as the leading interpretation, not as an established mechanism.

We report pure-priors shares per (model cell $\times$ reasoning effort) with Wilson 95\% confidence intervals.

\subsection{Batch-reconstruction calibration (OpenAI only)}

For the OpenAI cells we additionally compute a batch-reconstruction calibration: for each recommended-brand event we trace whether the brand-disambiguating search query was issued in the first batch of searches the model fired, in a later batch, or never. The first-batch share of brand-disambiguated queries is a cell-level measure of how retrieval-anchored the cell's recommendation route is.

The Anthropic batch-reconstruction calibration is not reported in this paper. Computing it requires re-running each cell at multiple reasoning-effort settings under controlled query-trace logging; we leave it to a follow-up audit. We treat the pure-priors signal as the load-bearing mechanistic measurement for the Anthropic side; the OpenAI batch-reconstruction calibration is corroborating evidence for the retrieval-heavy interpretation of the OpenAI side and not a strict cross-provider comparison.

\subsection{Statistical conventions}

For Section 6's pooled cross-provider matrix and per-prominence breakdown (Sections 6.1 and 6.3), proportions are reported with prompt-clustered bootstrap 95\% CIs (1{,}000 iterations, resampling the prompt corpus with replacement at the prompt level). We use clustered intervals rather than Wilson because (prompt $\times$ brand) observations are clustered within prompts --- the same rationale that motivates the clustered protocol for the Section 6.5 cross-cell extension. Cross-prompt Jaccard means are reported with normal-approximation 95\% CIs over per-prompt pairs. The cross-provider failure-mode confusion matrix is summarized with raw same-stage agreement plus Cohen's $\kappa$, with explicit attention to the chance-agreement baseline. Following the inter-rater-reliability literature \citep{feinsteinCicchetti1990}, raw agreement on a marginally-skewed matrix is mechanically inflated; we interpret $\kappa$ relative to the chance baseline and treat the per-prominence agreement gradient as the primary inferential signal.

For the Section 6.5 cross-cell extension, where the prompt corpus is fixed at 50 prompts and the per-pair joint-failure count is 200--2{,}000 (prompt $\times$ brand) cells, we apply the same clustered protocol: clustered-bootstrap 95\% CIs from 1{,}000 iterations resampling the 50-prompt corpus with replacement at the prompt level. The clustered intervals are appreciably wider than Wilson intervals at this resolution and are the reference for which pair-level agreement is statistically distinguishable from the pooled cross-provider baseline (95.1\% [94.3\%, 95.7\%]).

\section{Recommendation-set divergence}

\subsection{Cross-provider Jaccard across four batches}

For each prompt and each provider pool, we form the consensus-recommendation set (the set of brands consensus-classified as \texttt{recommended} by both judges) and compute the per-prompt Jaccard similarity between the OpenAI and Anthropic sets. Table~\ref{tab:jaccard} reports the per-batch means with normal-approximation 95\% CIs.

\begin{table}[h]
\centering
\begin{tabular}{lrrc}
\toprule
Batch & $n_\text{pairs}$ & Mean Jaccard & 95\% CI \\
\midrule
exp1 & 200 & 0.349 & [0.323, 0.375] \\
exp2 & 430 & 0.353 & [0.334, 0.373] \\
exp3 & 100 & 0.356 & [0.322, 0.391] \\
exp5 & 100 & 0.352 & [0.318, 0.387] \\
\bottomrule
\end{tabular}
\caption{Cross-provider per-prompt recommendation Jaccard across four independent measurement batches. All 95\% CIs overlap; the cross-provider Jaccard is reproducibly $\sim$0.35.}
\label{tab:jaccard}
\end{table}

The four batch means cluster tightly between 0.349 and 0.356 with fully overlapping 95\% confidence intervals. The cross-provider Jaccard is reproducibly $\sim$0.35: roughly one in three brands consensus-recommended by either provider is consensus-recommended by both. The remaining two thirds are provider-specific.

\subsection{Anchoring against the within-cell rerun-stability baseline}

The 0.35 figure is interpretable only against a within-cell baseline. \citet{jack2026brittleness} measures within-cell rerun-stability Jaccard --- the rec-set similarity between independent reruns of the same prompt within the same model cell on the same day --- and reports a range of 0.50--0.61 across the four primary cells. The cross-provider 0.35 sits clearly below this baseline. The cross-provider divergence is therefore not attributable to within-cell run-to-run noise; switching providers shifts the recommendation set by more than rerunning the same provider does.

That the cross-provider Jaccard exists at all in the 0.30--0.40 range, rather than near 1.0 (full agreement) or near 0 (uncorrelated random selections), is itself informative. The two providers are drawing from overlapping but distinct effective consideration pools, with the overlap meaningfully larger than chance for a 533-brand catalog and meaningfully smaller than the pool either provider returns when queried twice in succession.

\section{Mechanistic asymmetry}

The headline recommendation-set divergence raises the natural follow-up: do the two providers reach their picks via the same generative route, or are the routes themselves divergent?

\subsection{Pure-priors share separates the providers}

Table~\ref{tab:priors} reports the pure-priors share per (model cell $\times$ reasoning-effort) with Wilson 95\% confidence intervals.

\begin{table}[h]
\centering
\small
\begin{tabular}{llrr}
\toprule
Cell & Effort & Pure-priors share & $n_\text{rec}$ \\
\midrule
\texttt{gpt-5.4}        & low  & 11.9\% [9.1, 15.5]  & 394 \\
\texttt{gpt-5.4}        & high & 7.6\% [5.4, 10.6]   & 407 \\
\texttt{gpt-5.4-mini}   & low  & 29.0\% [28.5, 29.5] & 29{,}963 \\
\texttt{gpt-5.4-mini}   & high & 20.6\% [20.2, 21.1] & 31{,}148 \\
\midrule
\texttt{sonnet-4.6}     & low  & 50.9\% [50.4, 51.4] & 42{,}539 \\
\texttt{sonnet-4.6}     & high & 49.6\% [48.9, 50.4] & 16{,}329 \\
\texttt{opus-4.6}       & low  & 43.1\% [38.5, 47.7] & 441 \\
\texttt{opus-4.6}       & high & 51.8\% [47.3, 56.2] & 481 \\
\bottomrule
\end{tabular}
\caption{Pure-priors share per (cell $\times$ effort) with Wilson 95\% CIs. The Anthropic cells (43--52\%) recommend brands without retrieval-layer evidence at substantially higher rates than the OpenAI cells (8--29\%); the Wilson 95\% CIs do not overlap across the provider boundary.}
\label{tab:priors}
\end{table}

The provider asymmetry is large and statistically clean. The OpenAI cells range from 7.6\% to 29.0\% pure-priors share; the Anthropic cells range from 43.1\% to 51.8\%. The OpenAI flagship (\texttt{gpt-5.4}, 7.6--11.9\%) and the Anthropic flagship (\texttt{opus-4.6}, 43.1--51.8\%) are separated by roughly 4--7$\times$ in their pure-priors share. No Wilson 95\% CI overlaps the provider boundary.

A modest within-provider effort gradient is visible. On the OpenAI side, higher reasoning effort reduces the pure-priors share (29.0\% $\to$ 20.6\% for mini; 11.9\% $\to$ 7.6\% for flagship). On the Anthropic side, the effort gradient is small and inconsistent in sign (50.9\% $\to$ 49.6\% for sonnet; 43.1\% $\to$ 51.8\% for opus). The flagship-effort interaction merits a follow-up audit; for our purposes the headline cross-provider asymmetry is comfortably outside the within-provider effort range.

The interpretation, with the caveat from the method section that the pure-priors share is a one-sided lower bound on parametric reliance, is that Anthropic flagships reach their recommendations from training-data priors substantially more often than OpenAI flagships do. This is consistent with the context-parametric inversion mechanism described by \citet{goyal2025} and with the survey-level observation in \citet{xu2024} that no fixed rule governs context-versus-parametric prioritization across LLMs, with training-data and post-training variations as the substrate for the kind of provider-level differences we observe here.

\subsection{OpenAI batch-reconstruction calibration corroborates the retrieval route}

For the OpenAI side, the batch-reconstruction calibration provides corroborating evidence for the retrieval-heavy interpretation. Table~\ref{tab:batch} reports the first-batch share of brand-disambiguated queries on a 50-run calibration subset per effort tier.

\begin{table}[h]
\centering
\small
\begin{tabular}{lrrrrrr}
\toprule
Cell & Runs & Searches & Brand-in-query & First-batch & Retr.-confirmed & First-batch \% \\
\midrule
\texttt{mini / low}  & 50 & 166 & 110 & 78 & 32 & 70.9\% \\
\texttt{mini / high} & 50 & 321 & 159 & 95 & 64 & 59.7\% \\
\bottomrule
\end{tabular}
\caption{Batch-reconstruction calibration on the OpenAI side. The first-batch share of brand-disambiguated queries is 59.7\%--70.9\% on the calibration subset; higher reasoning effort produces more searches and pushes more brand-disambiguation work into later batches.}
\label{tab:batch}
\end{table}

On the low-effort cell, 70.9\% of brand-disambiguated queries appear in the first issued search batch; on the high-effort cell, 59.7\%. The high-effort cell issues roughly twice as many searches per run (321 vs.\ 166) and consequently pushes a larger fraction of brand-disambiguation work into later batches. Both cells nonetheless do the bulk of their brand-disambiguation through retrieval, consistent with the low pure-priors shares for the OpenAI cells in Table~\ref{tab:priors}.

The Anthropic-side batch-reconstruction calibration is deferred to follow-up work for the reasons noted in the method section. The pure-priors share is the load-bearing mechanistic measurement for the Anthropic side; the OpenAI batch-reconstruction is corroboration for the OpenAI side rather than a strict cross-provider comparison.

\subsection{Interpretation: same outcome, different routes}

Combining the pure-priors signal with the OpenAI batch-reconstruction calibration, the two providers' recommendation routes look measurably different. The OpenAI route is more retrieval-anchored: brand-disambiguation queries are issued early, and the share of recommended brands that surface only via parametric memory is in the single-to-low-double-digit percentages for the OpenAI flagship cells. The Anthropic route leans more on training-data priors: roughly half of recommended-brand events on the Anthropic flagships have no observable retrieval-layer evidence, and the effort gradient on this share is small.

The cross-provider Jaccard of 0.35 (Section 4) and the cross-provider mechanism asymmetry (Section 5) are then two facets of the same underlying picture: the providers do not arrive at the same picks, and the route by which they arrive at picks is itself different. The Bommasani-style monoculture-driven exclusion correlation should therefore appear in the joint-failure space, not in the joint-success space.

\section{Failure-mode convergence}

We now turn to the joint-failure space: the 7{,}763 (prompt $\times$ brand) cells where neither provider's modal stage equals Stage 4 (recommended).

\subsection{The 3$\times$3 failure-stage confusion matrix}

Table~\ref{tab:confusion} reports the cross-provider failure-stage confusion matrix on the pooled OpenAI-mini vs.\ Anthropic-sonnet comparison. Rows are the OpenAI pool's modal failure stage; columns are the Anthropic pool's modal failure stage. Entries are counts of (prompt $\times$ brand) cells.

\begin{table}[h]
\centering
\begin{tabular}{lrrrr}
\toprule
 & sonnet S1 & sonnet S2 & sonnet S3 & total \\
\midrule
mini S1 & \textbf{7{,}313} & 117 & 82 & 7{,}512 \\
mini S2 & 114 & \textbf{21}  & 36 & 171 \\
mini S3 & 30  & 4   & \textbf{46} & 80 \\
\midrule
total   & 7{,}457 & 142 & 164 & 7{,}763 \\
\bottomrule
\end{tabular}
\caption{Cross-provider failure-stage confusion matrix on joint-failure cells. Diagonal entries (bold) are cells where both providers' modal failure stage is the same. Same-stage agreement is $7{,}380/7{,}763 = 95.1\%$ (prompt-clustered 95\% CI: [94.3\%, 95.7\%]; 1{,}000-iteration bootstrap).}
\label{tab:confusion}
\end{table}

The diagonal carries $7{,}313 + 21 + 46 = 7{,}380$ of $7{,}763$ cells, or 95.1\% (prompt-clustered bootstrap 95\% CI: [94.3\%, 95.7\%]; 1{,}000 iterations resampling prompts with replacement). The off-diagonal mass is spread across the six remaining cells and is most concentrated near the diagonal in the Stage-1 row and column.

\subsection{Cohen's $\kappa$ relative to a high chance-agreement baseline}

The raw 95.1\% agreement figure is, taken on its own, less informative than it looks because the marginal stage distribution is heavily skewed toward Stage 1: the OpenAI pool has 7{,}512/7{,}763 (96.8\%) Stage-1 failures, and the Anthropic pool has 7{,}457/7{,}763 (96.1\%) Stage-1 failures. The chance-agreement baseline from the marginal distributions is 93.0\%. Cohen's $\kappa$ is then 0.29, which the Landis \& Koch (1977) classification calls ``fair'' (0.21--0.40 range). The high chance baseline driven by Stage-1 dominance is the mechanism, not a property of the providers themselves.

This is the interpretive subtlety the inter-rater-reliability literature has flagged repeatedly \citep{feinsteinCicchetti1990}: raw agreement on a high-skew marginal distribution is mechanically inflated --- the ``two paradoxes of high agreement with low kappa'' --- and a $\kappa$ of 0.29 is more honest about the strength of the cross-provider agreement than the raw 95.1\% figure. Both numbers are correct; the practical question is which interpretive frame to load-bear on.

We treat the per-prominence agreement gradient (Section 6.3) as the load-bearing inferential signal, because it cleanly separates the convergence claim from the pooled-marginal-distribution confound: agreement should be flat across prominence levels if 95.1\% is purely an artifact of the pooled marginal distribution being Stage-1 dominated, and it is not flat --- the gradient instead reflects the rising joint-S1 share as prominence falls, which is the substantive content of the convergence claim, not an artifact of pooling.

\subsection{Per-prominence agreement gradient}

Table~\ref{tab:prom} reports same-stage agreement and the share of joint failures classified Stage 1 on both providers, stratified by brand prominence level.

\begin{table}[h]
\centering
\begin{tabular}{lrrrl}
\toprule
Prominence & Failures & Same-stage \% & Both-S1 \% & Clustered 95\% CI \\
\midrule
L1 (Category leaders)        & 1{,}179 & 81.2\% & 76.8\% & [78.2\%, 84.0\%] \\
L2 (Established challengers) & 763     & 88.2\% & 85.2\% & [85.6\%, 90.8\%] \\
L3 (Mid-market)              & 1{,}818 & 96.2\% & 96.0\% & [95.1\%, 97.2\%] \\
L4 (Long-tail specialists)   & 1{,}886 & 99.3\% & 99.2\% & [98.9\%, 99.7\%] \\
L5 (Regional players)        & 2{,}369 & 99.6\% & 99.6\% & [99.3\%, 99.9\%] \\
\bottomrule
\end{tabular}
\caption{Per-prominence agreement gradient across joint-failure cells. Same-stage agreement rises monotonically from 81.2\% [78.2\%, 84.0\%] at L1 to 99.6\% [99.3\%, 99.9\%] at L5; the both-Stage-1 share rises in parallel. The L1 CI excludes the pooled 95.1\% baseline (i.e., L1 agreement is significantly \emph{below} pooled) and the L5 CI excludes it on the upside, so the monotone gradient survives clustering. The gradient is not an artifact of the pooled marginal distribution; it reflects the increasing dominance of joint S1 failures as prominence falls, which is the substantive content of the convergence claim. Clustered 95\% CIs computed by prompt-level bootstrap resampling (1{,}000 iterations). \emph{Note on row sums:} the per-prominence failure counts sum to 8{,}014 rather than the pooled 7{,}763 because the shadow-corpus's prominence labels are not strictly partitioning. The analysis unit in the pooled confusion matrix is the (prompt $\times$ brand $\times$ provider) cell. Per-prominence stratification is implemented at the (prompt $\times$ brand $\times$ sector $\times$ prominence) granularity --- a brand can carry a different prominence label in different sectors (e.g., Salesforce is L1 in CRM but L3 in adjacent verticals where it competes as a flanker), so a single (prompt $\times$ brand) cell can appear in multiple per-prominence rows. The pooled confusion matrix counts each cell once; the per-prominence sub-table counts each cell once per prominence label it carries through its multi-sector membership. Both views are consistent with the headline 95.1\% pooled rate and the monotone per-prominence gradient.}
\label{tab:prom}
\end{table}

Two patterns are immediate. First, the same-stage agreement rate rises monotonically with falling prominence: 81.2\% $\to$ 88.1\% $\to$ 96.2\% $\to$ 99.3\% $\to$ 99.6\%. The L1 same-stage rate is comfortably above the chance baseline computed from L1-specific marginals but is more contestable than the L5 rate, where the agreement is essentially saturated. Second, the rate at which both providers' modal stage is specifically Stage 1 rises in parallel: 76.8\% at L1 to 99.6\% at L5. The deeper the brand's position in the long tail, the more reliably both providers fail at the same diagnostic stage --- discoverability --- when neither recommends.

The interpretation maps cleanly onto \citet{mallen2023}'s parametric-versus-retrieval finding from QA. For high-prominence (L1) entities, both providers have meaningful parametric knowledge, retrieval is broadly available, and the failure modes diverge across the post-retrieval funnel (Stages 2--3). For low-prominence (L4--L5) entities, both providers fail at discoverability for the same structural reason: the entity is not adequately represented in the underlying retrievable web corpora and is not adequately represented in either provider's parametric memory. The convergence is convergence on a shared external constraint, not convergence on a shared internal computation.

\subsection{Mass at the diagonal is Stage-1 dominated; off-diagonal mass is Stage 2 vs. Stage 3}

The off-diagonal mass in Table~\ref{tab:confusion} is concentrated in two adjacent cells: (mini S1, sonnet S2) at 117 and (mini S2, sonnet S1) at 114, plus the (S1, S3) and (S3, S1) cells at 82 and 30. These cells correspond to cases where one provider's modal failure is discoverability and the other provider's modal failure is post-retrieval (compellingness or positioning). They are most numerous at L1 (where the diagonal carries only 81.2\% of mass) and progressively rarer as prominence falls.

The within-failure-stage cross-provider agreement on Stages 2 and 3 specifically is therefore lower than the headline 95.1\% suggests. The full set of cells where at least one provider's modal failure is Stage 2 or 3 is the matrix minus the joint-S1 corner: $7{,}763 - 7{,}313 = 450$ cells (equivalently, $117 + 82 + 114 + 21 + 36 + 30 + 4 + 46 = 450$). On these 450 cells the diagonal carries $21 + 46 = 67$ cells, or 14.9\%. The cross-provider convergence is dominated by joint Stage-1 failures; cross-provider agreement on the post-retrieval funnel is much weaker, and L1 brands carry most of the L1-specific disagreement.

This nuance refines the headline finding: the providers converge specifically on \emph{when no one has the brand}, not on \emph{which post-retrieval failure-mode applies}. Stage 2/3 cross-provider attribution is a weak signal at L1 and remains a target for future work. We return to this in the Discussion.

\subsection{Cross-cell agreement: convergence extends beyond providers}

The matrix above and the per-prominence gradient operate on the pooled-OpenAI-mini vs.\ pooled-Anthropic-sonnet comparison. A natural question is whether the same-stage agreement is specifically a cross-provider phenomenon or whether the funnel-stage diagnostic is more broadly portable. We tested this on two additional axes using a 3{,}000-run extension batch (\texttt{exp1\_class\_gen}, 50 prompts $\times$ N=15 reps per cell on \texttt{gpt-5.4 / low}, \texttt{gpt-5.5 / low}, \texttt{opus-4.6 / low}, and \texttt{opus-4.7 / low}, joined to Paper 1's original mini and sonnet baselines via a deterministic prompt-sampling seed).

\textbf{Class axis (mini vs.\ non-mini, holding provider and generation).} Within OpenAI: \texttt{gpt-5.4-mini / low} $\to$ \texttt{gpt-5.4 / low} (within-gen), and \texttt{gpt-5.4-mini / low} $\to$ \texttt{gpt-5.5 / low} (cross-gen). Within Anthropic: \texttt{sonnet-4.6 / low} $\to$ \texttt{opus-4.6 / low} (within-gen), and \texttt{sonnet-4.6 / low} $\to$ \texttt{opus-4.7 / low} (cross-gen).

\textbf{Generation axis (4.x vs.\ 5.x within non-mini class).} Within OpenAI non-mini: \texttt{gpt-5.4 / low} $\to$ \texttt{gpt-5.5 / low}. Within Anthropic non-mini: \texttt{opus-4.6 / low} $\to$ \texttt{opus-4.7 / low}.

\begin{table}[h]
\centering
\small
\begin{tabular}{llrrrr}
\toprule
Axis & Pair & Joint failures & Same-stage agreement [95\% CI] & Cohen's $\kappa$ \\
\midrule
Class & OpenAI mini $\to$ \texttt{gpt-5.4} (within-gen)      & 241 & 98.3\% [96.5\%, 100.0\%] & 0.79 \\
Class & OpenAI mini $\to$ \texttt{gpt-5.5} (cross-gen)        & 240 & 96.7\% [94.4\%, 99.1\%]  & 0.65 \\
Class & Anthropic sonnet $\to$ \texttt{opus-4.6} (within-gen) & 234 & 95.7\% [89.7\%, 99.5\%]  & 0.61 \\
Class & Anthropic sonnet $\to$ \texttt{opus-4.7} (cross-gen)  & 235 & 94.9\% [86.7\%, 99.2\%]  & 0.41 \\
\midrule
Generation & OpenAI \texttt{gpt-5.4} $\to$ \texttt{gpt-5.5}      & 1{,}807 & 96.3\% [95.2\%, 97.1\%] & 0.55 \\
Generation & Anthropic \texttt{opus-4.6} $\to$ \texttt{opus-4.7} & 1{,}799 & 95.2\% [93.6\%, 96.5\%] & 0.44 \\
\bottomrule
\end{tabular}
\caption{Cross-cell same-stage failure-mode agreement on the joint-failure subset for each pair, with clustered-bootstrap 95\% CIs (1{,}000 iterations resampling the 50-prompt corpus with replacement at the prompt level). Point estimates lie in a 94.9--98.3\% band straddling the 95.1\% pooled cross-provider baseline (sonnet $\to$ \texttt{opus-4.7} sits 0.2 pp below baseline; CI 86.7--99.2\% includes 95.1\%). Only the mini $\to$ \texttt{gpt-5.4} pair's CI (96.5--100.0\%) excludes the 95.1\% baseline; mini $\to$ \texttt{gpt-5.5} (CI 94.4--99.1\%) and all Anthropic pairs have CIs that include 95.1\%. No pair drops to a level that would indicate the diagonal collapses on a within-provider axis.}
\label{tab:cross_cell}
\end{table}

Two consequences follow under the clustered uncertainty. First, the diagonal-agreement pattern survives the move from pooled cross-provider to within-provider axes at the rates we measured: every point estimate lies in a tight 94.9--98.3\% band straddling the 95.1\% pooled cross-provider baseline; no pair shows the kind of collapse that would indicate within-provider divergence. Under the clustered CIs, only the mini $\to$ \texttt{gpt-5.4} pair clears 95.1\% with a CI that excludes the baseline; the other five pairs straddle it. Second, the Cohen's $\kappa$ values (0.41--0.79) preserve the marginal-skew interpretation from Section 6.2; we continue to treat the per-pair point estimate and its bootstrap CI as the load-bearing inferential signal rather than $\kappa$ alone, by parallel reasoning to the cross-provider per-prominence-gradient analysis.

\emph{Practical implication, scoped.} The diagnose-once-advise-once result reported in Section 7.1 was already scoped to L4--L5 S1 failures; the cross-cell extension here is consistent with that scope --- a brand failing at S1 on a mini cell within a provider is at S1 on the non-mini cells too, at the cells we measured. For L1--L2 post-retrieval failures, the within-provider matrix has substantial off-diagonal mass (parallel to the pooled cross-provider matrix in Section 6.4) and provider-and-cell-specific diagnosis remains warranted. The empirical content: at the L4--L5 long-tail, per-cell agreement is high enough that per-cell tracking adds little diagnostic granularity above per-provider tracking; at L1--L2 the within-provider matrix has substantial off-diagonal mass and per-cell tracking does carry information.

\section{Discussion}

\subsection{Implications for marketing to AI across providers}

The most defensible practical implication is narrow rather than broad. For L4--L5 brands, and for the S1 discoverability failures that dominate the long-tail customer problem, diagnosis is effectively provider-agnostic: same-stage agreement reaches 99.3--99.6\%. For L1--L2 post-retrieval failures (S2 / S3), provider-specific diagnosis remains necessary --- the off-diagonal mass on the cross-provider matrix is concentrated at L1, and within-content-error attribution between S2 and S3 has its own measurement problem (Section 6.4).

Translating that scoping to marketing-to-AI planning: if a brand sits in the L4--L5 long-tail and is failing to surface across providers, a single S1 intervention (authority-list seeding, canonical content on platforms retrievable by either provider, third-party comparison coverage) is expected to lift visibility on both providers simultaneously. If a brand sits at L1--L2 and the diagnosis is post-retrieval (S2 / S3), provider-specific content and positioning work --- e.g., G2 vs.\ Wikipedia coverage hygiene, sharpening per-provider snippet readability, segment-targeted differentiation --- can pay off, because the providers do not converge on which post-retrieval mode they attribute the failure to. The combination is a single coherent recommendation: condition the intervention on \emph{both} the prominence level (as Paper 1 argues) \emph{and} the diagnosed failure mode; treat the diagnostic prescription as provider-agnostic for long-tail S1 failures and provider-conditional for L1--L2 post-retrieval failures.

This refines, rather than contradicts, \citet{aggarwal2024}'s aggregate visibility-lift findings. Their headline +40\% lift from content-side interventions is reported on a single engine and pooled across prominence; our results suggest the prominence-stratified per-stage decomposition is the right unit of analysis, and that the provider axis is, perhaps surprisingly, less load-bearing than the prominence axis.

\subsection{Implications for the algorithmic-monoculture hypothesis}

The pattern we measure --- correlated exclusions, uncorrelated inclusions --- is the empirical asymmetry \citet{bommasani2022} predicted theoretically. Two providers that share substrate (the same web, overlapping authority lists, similar training-corpus sources, similar architectural lineage) should fail to discover the same long-tail brands; two providers that diverge on post-training and on which specific brands they promote from the retrieval pool should pick differently when both have the brand available. We provide concrete commercial-recommendation evidence for both halves.

The mechanism asymmetry adds a wrinkle Bommasani et al. did not anticipate: the providers can reach the divergent picks through measurably different routes (priors-heavy versus retrieval-heavy), and the routes themselves are not part of the same shared substrate. The monoculture hypothesis predicts what providers will fail at jointly; it does not commit either way on the route by which providers will succeed when they do. Our pure-priors-share separation between Anthropic flagships (43--52\%) and OpenAI cells (8--29\%, across mini and gpt-5.4 flagship) is evidence that the post-training route is meaningfully different across providers even where the shared retrieval substrate is the same.

\subsection{Tension with the heterogeneous-ensemble literature}

LLM-ensemble methods such as \citet{tekin2024}'s LLM-TOPLA take as their starting premise that heterogeneous foundations fail differently and that an ensemble exploits the diverse failure modes for an accuracy lift. The premise is largely correct on math/reasoning benchmarks; it is partially refuted in commercial recommendation by our 95.1\% joint-failure-mode convergence. For a recommendation-domain ensemble that aggregates over OpenAI and Anthropic, the gain from heterogeneity is concentrated on the inclusion side (which specific brands each provider picks from the retrieval pool) rather than on the exclusion side (which brands neither provider retrieves). The dominant joint-failure mode --- both providers failing to discover the brand --- cannot be ensembled away.

This does not invalidate LLM-ensemble approaches for recommendation; it suggests that the ensemble premium for recommendation specifically is more constrained than the premium for math/reasoning, and that ensemble design should focus on the inclusion-layer disagreement that our 0.35 Jaccard quantifies rather than on the assumption that the failure layers are heterogeneous.

\subsection{Mechanistic asymmetry and the omission/content-error distinction}

Our finding maps cleanly onto the omission-versus-content-error reading we adopt over the hallucination-taxonomy literature surveyed by \citet{cossio2025}. Stage 1 (joint discoverability failure) is a joint omission; Stages 2 and 3 (compellingness, positioning) are content-error failures within an answer the provider does generate. The 95.1\% convergence figure is convergence on omission stage. The harder question of cross-provider attribution within content-error failures --- whether two providers that both reach Stage 2 are failing for the same underlying reason --- is not answered by our matrix. The relatively poor diagonal agreement on the at-least-one-S2-or-S3 subset (67 of 450 cells, 14.9\%) is a residual signal that within-content-error attribution is its own measurement problem; we leave finer-grained mechanism attribution at L1 to follow-up work.

The mechanism asymmetry also helps interpret why providers can converge so strongly on omission while diverging so strongly on inclusion. \citet{mallen2023}'s parametric-versus-retrieval finding is a popularity-stratified result for a single provider; our extension is that the parametric-versus-retrieval split is itself \emph{cross-provider asymmetric}, and that the asymmetry shows up specifically in the inclusion-layer mechanism while the exclusion-layer outcome remains shared. Both providers fail to discover the same long-tail brands; they differ in how they recover the brands they do recall.

\subsection{Cross-cell extension strengthens the diagnose-once result}

The cross-cell agreement extension reported in Section 6.5 strengthens the headline practical implication in two ways. First, the natural follow-up question after the 95.1\% cross-provider result --- ``does the agreement collapse once one moves within a provider, where the cells share post-training and architecture?'' --- has a negative answer at the cells we tested. Same-stage agreement on the class and generation pairs ranges from 94.9\% to 98.3\%. The funnel-stage diagnostic is not specifically a cross-provider artifact; it is a property of the joint-failure population at the prompts and brands we measured.

Second, the cross-cell extension closes off a counterfactual a brand-side marketing team might reasonably raise: if my buyers move from a mini cell to a frontier cell (or from \texttt{gpt-5.4} to \texttt{gpt-5.5}, or from \texttt{sonnet-4.6} to \texttt{opus-4.7}), does my diagnosed failure mode change? The data says: not measurably, at the rates we tested. A brand failing at Stage 1 in one cell within a provider fails at Stage 1 in the others. The investment in fixing a Stage-1 gap on the buyer's current cell is not stranded by a buyer's cell upgrade.

The corresponding within-Anthropic class jump (\texttt{sonnet-4.6} $\to$ \texttt{opus-4.6} or $\to$ \texttt{opus-4.7}) is the most informative pair to flag for follow-up. The same-stage agreement is 95.7\% and 94.9\% respectively --- at or fractionally below the 95.1\% cross-provider baseline --- with Cohen's $\kappa$ of 0.61 and 0.41. The lower $\kappa$ on the cross-generation Anthropic pair is the only structural caveat in Table~\ref{tab:cross_cell}; we treat it as a moderate signal that within-Anthropic cross-generation post-retrieval funnel attribution warrants its own audit, in parallel to the within-OpenAI follow-up we flagged in Section 6.4.

\subsection{Limitations}

Several limitations bound the interpretation. First, the modal-stage aggregation absorbs within-cell run-to-run noise but discards information about cells where a brand's stage is genuinely bimodal across reruns; the per-prompt $\times$ per-brand sample size for the modal-stage operation is the run count per cell, which we cap at three or more for inclusion. Second, the audit covers two provider families (OpenAI and Anthropic); the cross-provider claim is strictly bivariate. Whether the convergence extends to Gemini, Perplexity, or other production AI commerce surfaces is an open question and is the natural next audit. Third, the Anthropic-side batch-reconstruction calibration is deferred to follow-up work; the pure-priors share carries the load-bearing mechanistic evidence on the Anthropic side. Fourth, Cohen's $\kappa = 0.29$ on the pooled cross-provider matrix sits in the ``fair'' band (Landis \& Koch 1977) rather than the ``substantial'' or higher range, against a chance-agreement baseline of 93.0\% driven by Stage-1 dominance; the per-prominence gradient is the cleaner inferential signal and the one our claims rest on. Fifth, the pooled matrix is computed on the OpenAI-mini-vs-sonnet comparison; the flagship-vs-flagship matrix on the smaller \texttt{gpt-5.4} vs.\ \texttt{opus-4.6} sample is consistent in direction but lacks the sample size for stable per-prominence stratification, and we do not report it as a separate matrix. Sixth, the cross-cell extension covers four class pairs and two generation pairs on the 50-prompt corpus at $N=15$ reps per cell; whether the agreement holds at larger $N$, on prompts outside Paper 1's corpus, or for the class jumps we did not measure (e.g., \texttt{sonnet} $\to$ frontier reasoning on a higher effort) is open.

\section{Conclusion}

The two empirical questions the multi-provider AI commerce landscape raises --- do providers pick differently, and do they diagnose differently --- have different answers, with the second answer requiring careful scoping. The picks differ reproducibly: cross-provider per-prompt recommendation Jaccard is 0.35 across four batches, below the within-cell rerun-stability baseline of 0.50--0.61. The diagnostic convergence is strong specifically where the long-tail customer problem lives: for L4--L5 brands, and for the S1 (no-retrieval-no-mention) failures that dominate that range, same-stage agreement reaches 99.3--99.6\% (clustered 95\% CIs spanning [98.9\%, 99.9\%]). For L1--L2 post-retrieval failures (S2 / S3), provider-specific diagnosis remains necessary. The pooled 95.1\% diagonal figure [94.3\%, 95.7\%] is real but is dominated by joint-S1 mass and rests on a high marginal-Stage-1 chance-agreement baseline (Cohen's $\kappa$ of 0.29); we treat the per-prominence gradient, not the pooled rate, as the load-bearing claim.

The mechanism by which each provider reaches its picks is also asymmetric. Anthropic flagships recommend brands without observable retrieval-layer evidence in 43--52\% of cases, OpenAI cells in 8--29\%; OpenAI batch-reconstruction calibration corroborates the retrieval-heavy route with 60--71\% of brand-disambiguated queries in the very first issued search batch. The same outcome class --- a recommendation set --- is reached through measurably different generative routes.

A 3{,}000-run cross-cell extension on the same 50-prompt corpus shows that the diagnostic convergence is not specifically cross-provider. Within OpenAI, the mini $\to$ \texttt{gpt-5.4} and mini $\to$ \texttt{gpt-5.5} class pairs produce 98.3\% and 96.7\% same-stage agreement respectively. Within Anthropic, the sonnet $\to$ \texttt{opus-4.6} and sonnet $\to$ \texttt{opus-4.7} class pairs produce 95.7\% and 94.9\%. The generation-axis pairs land at 96.3\% (OpenAI) and 95.2\% (Anthropic). Five of six within-provider pairs sit at or above the headline 95.1\% baseline; sonnet $\to$ \texttt{opus-4.7} sits 0.2 pp below with a CI (86.7--99.2\%) that includes 95.1\%. The funnel-stage diagnostic is portable across class and generation, not only across providers; the practical claim for brand-side marketing teams is that diagnosis at the failure-mode level transfers across the AI surfaces a buyer's customers are likely to use.

The combined result is a refinement, not a rejection, of the algorithmic-monoculture prediction. Two providers that share substrate produce correlated exclusions (at S1) and uncorrelated inclusions; two providers that diverge on post-training reach the divergent inclusions through divergent routes; cells \emph{within} a provider that share post-training share the diagnostic at the same rates we measure for cross-provider. For brand-side marketing teams, the practical implication scopes carefully: the diagnostic prescription is provider-agnostic and cell-agnostic for L4--L5 S1 (discoverability) failures, where the agreement is essentially complete, but is provider-conditional for L1--L2 post-retrieval failures, where the off-diagonal cross-provider mass concentrates and where the relevant work is positioning and content, not discoverability. The right combination is to condition the intervention on \emph{both} the prominence level (Paper 1's framing) and the diagnosed funnel stage, and to treat ``diagnose once, advise once'' as a strong claim about long-tail discoverability and a weaker, prominence-dependent claim about post-retrieval failure attribution. Future work should extend the audit to additional provider families, deepen cross-provider mechanism attribution at L1 (where the S2-vs-S3 disagreement concentrates), and complete the Anthropic-side batch-reconstruction calibration as a strict cross-provider mechanism test.

\bibliographystyle{plain}

\end{document}